\documentclass[aps,prl,twocolumn,superscriptaddress]{revtex4}

\usepackage{graphicx}    % include figure files
\usepackage{dcolumn}     % align table columns on decimal point
\usepackage{bm}          % bold math

\begin{document}

% Use the \preprint command to place your local institutional report
% number in the upper righthand corner of the title page in preprint mode.
% Multiple \preprint commands are allowed.
% Use the 'preprintnumbers' class option to override journal defaults
% to display numbers if necessary
%\preprint{}

%Title of paper
\title{Origin of spontaneous electric dipoles in homonuclear niobium clusters}

% repeat the \author .. \affiliation  etc. as needed
% \email, \thanks, \homepage, \altaffiliation all apply to the current
% author. Explanatory text should go in the []'s, actual e-mail
% address or url should go in the {}'s for \email and \homepage.
% Please use the appropriate macro foreach each type of information
% \affiliation command applies to all authors since the last
% \affiliation command. The \affiliation command should follow the
% other information
% \affiliation can be followed by \email, \homepage, \thanks as well.

\author{Kristopher E. Andersen}
\affiliation{Department of Physics, University of California,
  Davis, CA 95616-8677, USA}

\author{Vijay Kumar}
\affiliation{Dr. Vijay Kumar Foundation, 45 Bazaar Street,
  Chennai 600 078, India}
\affiliation{Institute for Materials Research, Tohoku
    University, 2-1-1 Katahira Aoba-ku, Sendai 980-8577, Japan}

\author{Yoshiyuki Kawazoe}
\affiliation{Institute for Materials Research, Tohoku
    University, 2-1-1 Katahira Aoba-ku, Sendai 980-8577, Japan}

\author{Warren E. Pickett}
\email{pickett@physics.ucdavis.edu}
\affiliation{Department of Physics, University of California,
  Davis, CA 95616-8677, USA}

%Collaboration name if desired (requires use of superscriptaddress
%option in \documentclass). \noaffiliation is required (may also be
%used with the \author command).
%\collaboration can be followed by \email, \homepage, \thanks as well.
%\collaboration{}
%\noaffiliation

\date{\today}

\begin{abstract}
Surprisingly large spontaneous electric dipole moments recently
observed in homonuclear niobium clusters below 100 K (Moro {\it et
al.} {\bf 300}, 1265 (2003)) are explained using first-principles
electronic structure calculations. The calculated moments for Nb$_n$
($n \leq 15$) closely follow the experimental data in which large
dipole moments are seen for $n$ = 11--14. We establish that the
dipoles are strongly correlated with the geometrical asymmetry of the
clusters. The magnitude of the dipole moment is roughly proportional
to the spread in the principal moments of inertia and its direction
tends to align along the axis of the largest principal moment. Charge
deformation densities reveal directional, partially covalent bonds
that enhance the formation of asymmetric geometries. Classical
simulations of the deflection of a cluster in a molecular beam reveal
that the electronic dipole may persist at higher temperatures, but is
masked by the rotational dynamics of the cluster.
\end{abstract}

% insert suggested PACS numbers in braces on next line
\pacs{}

% insert suggested keywords - APS authors don't need to do this
%\keywords{}

%\maketitle must follow title, authors, abstract, \pacs, and \keywords
\maketitle

Though the response of nanoparticles to an external electric field has
been actively studied for several years, the recent observation of
permanent electric dipoles on the order of several debye (D) in free
Nb$_n$ clusters ($n \sim $10--100) has introduced an entirely new and
unexpected dimension. That the intrinsic moment is only observed at
temperatures T($n \sim 10$) $\approx$ 100 K\cite{MXYdH03} has led to
speculation that the origin of large electric dipoles might be related
to superconductivity in bulk Nb (rather robust at T$_c$=9 K) and that
the dependence on temperature may signal a nanoscale ``ferroelectric''
transition. Permanent dipoles in mixed clusters and molecules,
including canonical polar molecules like H$_2$O and HF, are well
understood in terms of charge transfer between unlike atoms. However,
it is more challenging to envision what forces could drive the
underlying charge separation in homonuclear clusters of metal atoms.
This unexpected and nonintuitive dielectric behavior in Nb
clusters\cite{MXYdH03} has also attracted attention to the
conceptually related question of quantum dipoles in symmetric
clusters\cite{All04,AA03}.

A fundamental requirement for the occurrence of a permanent dipole is
the lack of a center of inversion. For a homonuclear system, however,
the issue is more complex because disproportionation into charged ions
is energetically disfavored. These questions are more perplexing in
clusters that are going to become metals (which support no charge
separation) as their size increases. The calculations we discuss
demonstrate that the origin of these large dipoles does not require a
new state of matter related to superconductivity as suggested in Ref.
\onlinecite{MXYdH03}, but is explained directly in terms of
directional partially covalent bonds that favor the formation of
clusters with low symmetry. The resulting charge separation, which
implies internal electric fields on the order of 10$^6$
V/cm\cite{MXYdH03}, is sustained by the chemical forces of the Nb
bonding. By performing classical simulations of the rotational
dynamics of the clusters in a molecular beam, we provide evidence that
thermal averaging, instead of a loss of intrinsic moment, accounts for
the observed temperature dependence.

The atomic and electronic structures of neutral Nb$_n$ clusters ($n
\leq 23$) have been extensively studied using first-principles
electronic structure methods\cite{GS93,GR96,GRA98,KK02}, but there has
been no theoretical study of their dielectric properties. In this
Letter, the electric dipole moments of Nb$_n$ clusters ($n \leq 15$)
are analyzed in terms of their asymmetry and charge deformation using
the electronic structure codes Gaussian98\cite{gaussian} and
Abinit\cite{abinit}. Calculations with Gaussian98 used
Stuttgart-Dresden effective core potentials\cite{AHD+90} and the
B3PW91\cite{PCV+92,PW92} hybrid exchange-correlation functional.
Calculations with Abinit\cite{abinit} used the planewave
(norm-conserving) pseudopotential method\footnote{The niobium
pseudopotential was generated using FHI98PP\cite{FS99} code and
included 4s and 4p semicore states. A cutoff energy of 90 Ry was used
to converge the planewave basis set. The cluster was placed within a
simple cubic supercell of 15 \AA. Only the $\Gamma$-point was used for
Brillouin zone integrations.} and PBE96\cite{PBE96} GGA functional.
The lowest energy structures as well as isomers close in energy
(within 25 meV/atom) obtained by Kumar and Kawazoe\cite{KK02} were
used, except for Nb$_{13}$, for which a new structure lower in energy
was found. Isomers of Nb$_n^+$ clusters have been observed in
spectroscopic\cite{KY90} and reactivity\cite{BSK+00} experiments.

\begin{figure}
\includegraphics[width=3in]{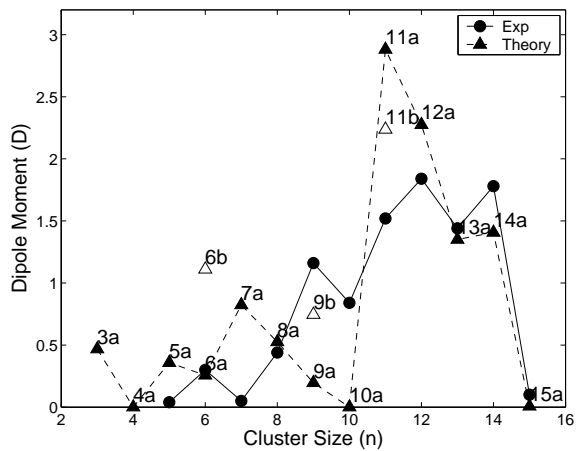}
\caption{\label{dipole}Comparison of the calculated dipole moments and
``low temperature'' (50 K) experimental data from FIG. 2 of Ref.
\onlinecite{MXYdH03} (1 D = 0.20819 e\AA). Higher energy structures
 are labeled in alphabetical order, following the notation of Ref.
\onlinecite{KK02}.}
\end{figure}

The calculated permanent electric dipoles (TABLE \ref{results}) are
compared in FIG. \ref{dipole} with the reported values\cite{MXYdH03}
for Nb clusters emitted after reaching thermal equilibrium with He gas
at 50 K. Considering the many uncertainties in presenting such a
comparison, the level of agreement is satisfying: moderate moments are
found for $3 \leq n \leq 9$, large moments for $11 \leq n \leq 14$,
and essentially zero moment for $n$ = 4, 10, and 15. The only notable
discrepancy is Nb$_{10}$ for which experiments show significant dipole
moment\footnote{In FIG. 1 we include the value of the dipole moment
given by Moro {\it et al.} \cite{MXYdH03} in the inset of their FIG.
2, although elsewhere they state Nb$_{10}$ is one of the clusters for
which ``the ferroelectric component is essentially absent.''} while
theoretically it is zero because of the symmetric nature of the lowest
energy structure. A general feature in FIG. 1 is the larger calculated
dipole moment in comparison to experiment. This difference may be due
to the neglect of rotational dynamics in the experimental
analysis\cite{MXYdH03}. Calculations on isomers, such as Nb$_{9b}$ and
Nb$_{11b}$, whose binding energies are lower by 1 and 9 meV/atom
respectively, show dipole moments closer to the experimental
data\cite{MXYdH03}, which may be an indication of isomers in the
experimental conditions. However, Nb$_{6a}$ is in good agreement with
experiment in comparison to Nb$_{6b}$, whose binding energy is also
only moderately lower (20 meV/atom).

\begin{table}
\caption{\label{results} Principal moments of inertia $I_1$--$I_3$
  ($M_{\text{Nb}} \text{\AA}^2$) and dipole moment $\mu$ (D) for Nb$_n$.}
\begin{ruledtabular}
\begin{tabular}{ddddd}
 n &   I_1   &   I_2   &   I_3   & \mu \\
 2 &   0.00  &   2.21  &   2.21  & 0.0 \\
 3 &   2.57  &   3.00  &   5.56  & 0.5 \\
 4 &   6.33  &   6.33  &   6.33  & 0.0 \\
 5 &   7.49  &  11.15  &  11.69  & 0.4 \\
 6 &  10.51  &  16.06  &  17.69  & 0.3 \\
 7 &  17.69  &  20.96  &  26.38  & 0.8 \\
 8 &  21.37  &  23.36  &  27.61  & 0.5 \\
 9 &  28.76  &  29.08  &  32.05  & 0.2 \\
10 &  32.54  &  38.03  &  38.03  & 0.0 \\
11 &  34.32  &  48.32  &  57.74  & 2.9 \\
12 &  43.05  &  55.11  &  64.06  & 2.3 \\
13 &  51.28  &  61.35  &  75.42  & 1.4 \\
14 &  61.29  &  66.71  &  81.68  & 1.4 \\
15 &  73.59  &  76.44  &  76.45  & 0.0 \\
\end{tabular}
\end{ruledtabular}
\end{table}

\begin{figure}
\includegraphics[width=3in]{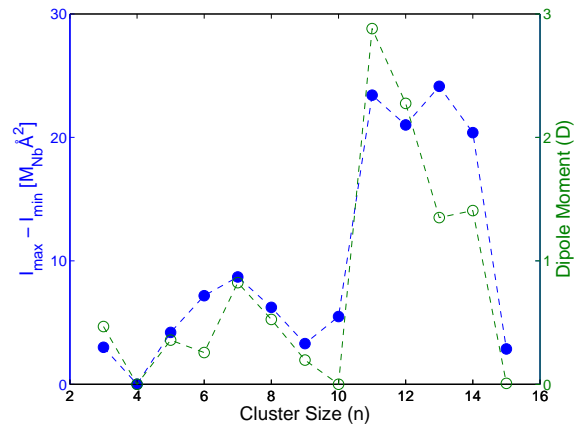}
\caption{\label{i-minmax} The strong correlation between the dipole
  moment (green, right axis) and asymmetry of the cluster, as
  quantified by the difference between the largest and smallest
  principal moments of inertia (blue, left axis).}
\end{figure}

For $n > 2$ only in the case of Nb$_4$, which is a regular
tetrahedron\cite{KK02}, does symmetry strictly forbid an electric
dipole. Other cases of vanishing dipole moment, such as Nb$_{10}$ and
Nb$_{15}$, are attributed to near symmetries. Nb$_{10}$ has an
approximate D$_{4d}$ symmetry, whereas Nb$_{15}$ has near inversion
symmetry with a cubic structure. These symmetries are reflected in the
cluster's principal moments of inertia $I_1$, $I_2$, and $I_3$ (TABLE
\ref{results}). Nb$_4$, for example, is a spherical top ($I_1 = I_2 =
I_3$), and Nb$_{10}$ and Nb$_{15}$ are symmetrical tops ($I_2 = I_3
\neq I_1$), although for Nb$_{15}$ the deviation from a spherical top
is quite small. The rest of the clusters studied are asymmetrical tops
($I_1 \neq I_2 \neq I_3$). One measure of the degree of asymmetry is
the spread $\Delta I = I_{\mbox{max}} - I_{\mbox{min}}$, where
$I_{\mbox{max}}$ ($I_{\mbox{min}}$) is the largest (smallest)
principal moment of inertia. Asymmetric clusters have correspondingly
large values of $\Delta I$. FIG. \ref{i-minmax} shows the strong
correlation between $\Delta I$ and the calculated electric dipole.
Since the principal moments of inertia map the dynamics of a rigid
body to that of an ellipsoid, $\Delta I$ can be interpreted as
quantifying the existence of a preferred structural axis in the
cluster. This interpretation is supported by the alignment of the
electric dipole. In cases where $\Delta I$ is large
(Nb$_{11}$--Nb$_{14}$), the direction of the dipole moment and the
principal axis corresponding to $I_{\mbox{max}}$ are close to
collinear. The 26$^\circ$ deviation from collinearity of Nb$_{12}$ is
the only exception to this otherwise near-perfect (within 5$^\circ$)
collinearity.

\begin{figure}
\includegraphics[width=2.25in,angle=-90]{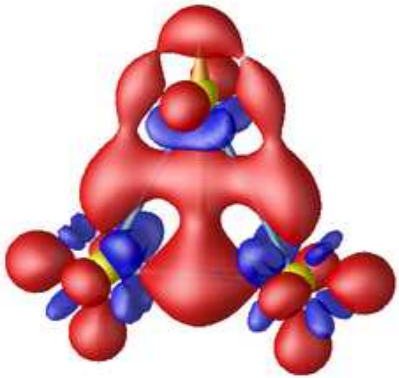}
\caption{\label{Nb3}Charge deformation of Nb$_3$ (0.5 D). Due to the
  mirror symmetry, the dipole moment (orange arrow) is restricted to
  lie on the symmetry axis.}
\end{figure}

To understand the origin of the electric dipole we consider the charge
deformation density due to bonding, that is, the difference between
the charge density of the cluster and that of the isostructural
spherical neutral atoms, for which the dipole moment vanishes. FIG.
\ref{Nb3} shows the charge deformation for the simple and illustrative
case of Nb$_3$\footnote{The dipole moment of Nb$_3$ is similar to that
of ozone O$_3$ (0.6 D).}, an isosceles triangle\cite{KK02} with a
dipole moment of 0.5 D. In FIG. \ref{Nb3}, two isosurfaces are shown.
The blue (red) surface corresponds to a negative (positive) isocontour
of value -2.0 e/\AA$^3$ (+2.0 e/\AA$^3$). Taken together, these
isosurfaces show how charge redistributes during the formation of
bonds. Two general features can be discerned. First is the formation
of a few specific covalent bonds as revealed by the charge build up
between ions, which reflects charge being pulled in from elsewhere in
the cluster. Second, some charge is pushed outward at each surface
site due to Coulomb repulsion.

\begin{figure}
\center
\includegraphics[width=2.25in]{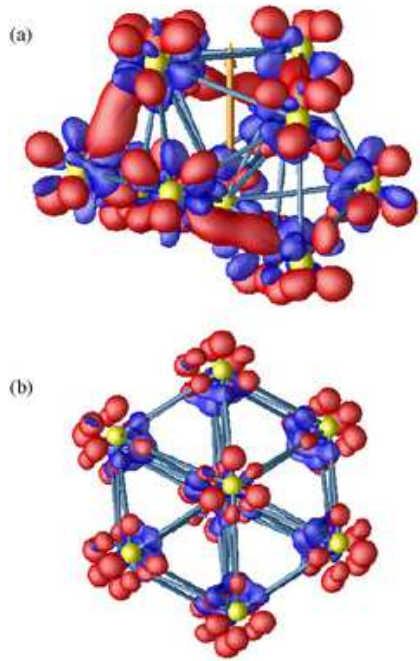}
\caption{\label{Nb12_Nb15}Charge deformation for (a) Nb$_{12}$ (2.3 D)
  and (b) Nb$_{15}$ (no dipole). The direction of the dipole moment is
  represented by the orange arrow. Note the formation of strong
  covalent bonding. Unlike systems with a large dipole moment, no
  strong asymmetric covalency is seen in Nb$_{15}$. Instead the charge
  deformation density is entirely intra-atomic.}
\end{figure}

FIG. \ref{Nb12_Nb15} shows the charge deformation for Nb$_{12}$ and
Nb$_{15}$, which are representative of clusters with large dipole
moment and no dipole moment respectively. The same color scheme and
isocontour values are used as in FIG. \ref{Nb3} and the general
features remain. Only in cases with a dipole moment, such as
Nb$_{12}$, is there a strong asymmetric character to the deformation
density. This reveals how strong directional bonding in isolated
regions of the cluster leads to the intrinsic charge separation giving
rise to asymmetry and the permanent dipole moment. We find however
that the regions in the cluster in which the charge deformation shows
this additional covalent character don't always coincide with the
shortest bonds. Instead, directional bonding between {\it d}-electrons
is important, driving the development of low-symmetry geometries.

An additional aspect of the experiments of Moro {\it et al.} is the
disappearance of the dipole ({\it i.e.}, a change in the deflection of
the cluster beam) when the clusters are formed at higher
temperatures\cite{MXYdH03}, which has been likened to a ferroelectric
transition. One possible mechanism for the disappearance of the
polarized electric state is the thermal excitation of vibrational
modes within the cluster, which should be directly related to the
temperature at which the clusters are formed. The charge response to
vibrations might disturb the intrinsic charge separation in the static
cluster at equilibrium. In addition, if the direction of the dipole
moment were to fluctuate, this charge separation, if still in
existence, could become unobservable if the time average of the dipole
moment $\langle\vec{\mu}\rangle$ were sufficiently reduced. Such
vibration-induced fluctuations of the dipole moment are known to be
important in weakly bound molecules\cite{CAR+02}. We suspect that such
vibrations are responsible for the zero dipole moment observed in
Nb$_5$ and Nb$_7$ which have distorted trigonal and pentagonal
bipyramid structures respectively. In the symmetric structure of
Nb$_7$, the highest occupied molecular orbital is doubly degenerate
but occupied with one electron, which leads to a Jahn-Teller
distortion. When formed and selected at a higher temperature, thermal
motion may move the cluster through several structures with differing
dipoles, the net effect of which would produce canceling deflections.

To quantify the influence of vibrations, we calculated the vibrational
spectrum for Nb$_{12}$, a cluster with a large permanent electric
dipole (2.3 D), and looked at the excitation of the 6 lowest energy
vibrational modes between 60--105 cm$^{-1}$. These are in the range of
those that would be excited during the experimentally observed
transition, which occurs at a temperature on the order of 100 K. The
atomic displacements associated with these vibrations are less than
0.1 \AA\ and are unable to significantly alter the charge density or
the permanent dipole, with relative changes of less than 3\% or 0.1 D
in magnitude and a few degrees in direction. This implies that another
mechanism, beyond harmonic vibrations, is needed to explain the
disappearance of the dipole moment.

\begin{figure}
\center
\includegraphics[width=3in]{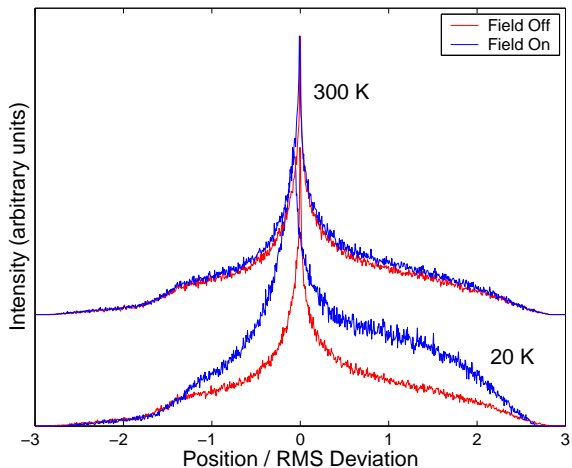}
\caption{\label{deflection}Classical deflection of a model Nb$_{12}$
cluster with a dipole moment $\mu$ = 2 D at 300 K (top) and 20 K
(bottom). At higher temperatures, the rotational dynamics masks the 
permanent dipole moment.}
\end{figure}

Classical simulations by Dugourd {\it et al.}\cite{DCL+01}, using the
thermal occupation of rotational states, were able to reproduce the
molecular beam deflection of TiC$_{60}$ clusters, but full temperature
dependence was not studied. FIG. \ref{deflection} shows the classical
deflection of a representative niobium cluster (Nb$_{12}$) with a
dipole moment of 2 D at 20 and 300 K when the homogenous electric
field is on and off. These profiles were obtained by numerically
evaluating the deflection\cite{DCL+01} of 10$^6$ random initial
configurations (orientation and rotational energy) of a cluster with
symmetric inertial moments I$_1$ = I$_2$ = 45 M$_{\text{Nb}}$ \AA$^2$
and I$_3$ = 65 M$_{\text{Nb}}$ \AA$^2$. The deflection due to the
induced dipole, which would shift each profile a constant distance,
has been ignored by setting the electric polarization $\alpha = 0$. In
addition, the broadening of the profile due to the cluster's reduced
translational velocity at lower temperature has been divided out.

The classical deflection profile FIG. \ref{deflection} and the
experimental profiles in Ref. \onlinecite{MXYdH03} agree
qualitatively. In particular, the asymmetric shape at lower
temperatures, which is an experimental signature of the permanent
moment\cite{MXYdH03}, is clearly seen in the profile at 20 K. At 300 K
the deflection profile is more symmetric, which has been interpreted
as a loss of the intrinsic moment\cite{MXYdH03}, but is actually
the effect of additional thermal averaging.

Finally we return to the question of (a lack of) symmetry that we
began with. Our calculations have traced the origin of large
spontaneous electric dipoles to the structural asymmetry of Nb$_n$
clusters that is enhanced by the directional bonding favored by {\it
d} electrons. These correlations can be sharpened. First, the large
asymmetric inertial moments observed in systems with large electric
dipole moment result from oblate, not prolate, cluster shapes. Second,
although the magnitude and direction of the electric dipole correlates
very strongly with an (asymmetric) imbalance of inertial moments, this
cannot be the whole story, since breaking reflection symmetry---of
which the inertial moment themselves provide no information---is
required for dipole formation. 

That the apparent temperature dependence of the electric dipole can be
explained classically implies a more rigorous treatment of rotational
dynamics\cite{HOB00} may be required to distinguish induced versus
permanent dipole effects in experiment accurately. However, because
the structure of the deflection profile is richer at low temperatures,
and depends directly on the inertial moments of the cluster, future
low-temperature molecular beam experiments have the potential to
measure structural parameters that are directly comparable to
theoretical predictions.

\begin{acknowledgments}
The authors would like to thank P. B. Allen for stimulating
discussions. KEA was supported by a DOE CSGF fellowship under grant
DE-FG02-97ER25308. WEP was supported by NSF grant DMR-0114818. VK
would like to acknowledge the hospitality at the IMR and the support
of the Center for Computational Materials Science, IMR-Tohoku
University for the use of the Hitachi SR8000/64 supercomputing
facilities. KEA and WEP would like to thank the NSF sponsored
NEAT-IGERT program for providing a stimulating environment.
\end{acknowledgments}

\bibliography{niobium}

\end{document}